\documentclass[12pt,letterpaper]{article}
\usepackage[utf8]{inputenc}

\usepackage{color}
\definecolor{darkblue}{rgb}{0.1,0.1,.7}
\usepackage[colorlinks, linkcolor=darkblue, citecolor=darkblue, urlcolor=darkblue, linktocpage]{hyperref}
\usepackage[]{amsmath}
\usepackage[]{graphicx}
\usepackage[]{latexsym}
\usepackage[utf8]{inputenc}
\usepackage{slashed,graphicx,color,amsmath,amssymb}
\usepackage[mathscr]{eucal}
\usepackage{mathrsfs}
\usepackage[margin=10pt,font=small,labelfont=bf]{caption}
\usepackage{amscd}
\usepackage{bm}
\usepackage{xcolor}
\usepackage{upgreek}
\usepackage[square, comma, sort&compress,numbers]{natbib}
\usepackage[all,cmtip]{xy}
\usepackage[margin=1.7in]{geometry}
\usepackage{cleveref}
\usepackage[color=cyan!30!white,linecolor=red,textsize=footnotesize]{todonotes}
\usepackage{soul}
\usepackage{tikz}
\usepackage{pgflibraryarrows}
\usepackage{pgflibrarysnakes}
\usepackage{pgfplots}

\pgfplotsset{compat=1.8}
\tikzset{elegant/.style={smooth,thick,samples=50,magenta}}
\numberwithin{equation}{section}
 \makeatletter
 \g@addto@macro\bfseries{\boldmath}

\makeatletter
\if@todonotes@disabled

\else

\fi
\makeatother

\usepackage{amsmath}
\begin{document}	
\vspace*{-.3in} \thispagestyle{empty}
\begin{flushright}
\end{flushright}
\vspace{.4in} {\Large
\begin{center}
{\bf On the symmetry of $T\bar{T}$ deformed CFT}
\end{center}}
\vspace{.2in}
\begin{center}
{ Miao He and Yi-hong Gao}
\\
\vspace{.3in}
\small{
\textit{School of Physical Sciences, University of Chinese Academy of Sciences, \\No.19A Yuquan Road, Beijing 100049, China}\\ \vspace{.1cm}
\vspace{.2cm}
\textit{CAS Key Laboratory of Theoretical Physics, Institute of Theoretical Physics, \\Chinese Academy of Sciences, Beijing 100190, China}
}\\ 
\vspace{.1cm}
\begingroup\ttfamily\small
hemiao@itp.ac.cn, gaoyh@itp.ac.cn
\endgroup 

\vspace{.1in}

\end{center}
\begin{abstract}
\normalsize{We propose a symmetry of $T\bar T$ deformed 2D CFT, which preserves the trace relation. The deformed conformal killing equation is obtained. Once we consider the background metric runs with the deformation parameter $\mu$, the deformation contributes an additional term in conformal killing equation, which plays the role of renormalization group flow of metric. The conformal symmetry coincides with the fixed point. On the gravity side, this deformed conformal killing equation can be described by a new boundary condition of AdS$_3$. In addition, based on the deformed conformal killing equation, we derive that the stress tensor of the deformed CFT equals to Brown-York's quasilocal stress tensor on a finite boundary with a counterterm. For a specific example, BTZ black hole, we get $T\bar T$ deformed conformal killing vectors and the associated conserved charges are also studied.}
\end{abstract}

\begin{center}
\end{center}

\newpage

\setcounter{page}{1}
\noindent\rule{\textwidth}{.1pt}\vspace{-1.2cm}
\begingroup
\hypersetup{linkcolor=black}
\tableofcontents
\endgroup
\noindent\rule{\textwidth}{.2pt}

\section{Introduction}
\label{sec: Intro}
In AdS/CFT correspondence, it is proposed that the AdS gravity is dual to a CFT living on the boundary at spatial infinity. It provides a powerful approach to explore quantum gravity and research strongly coupled system from gravity side. But whether the holographic correspondence can be extended to effective QFTs for which the UV behavior is not described by a CFT. Recently, Smirnov and Zamolodchikov discovered a class of exactly solvable irrelevant deformations of 2D QFTs called $T\bar T$ deformation~\cite{Smirnov2017}, which may provide a route to explore this question. The actions are defined by $T\bar T$ flow, and some properties have been studied in~\cite{Cavaglia2016}. The deformations generated by $T\bar T$ are solvable in a certain sense, even if the original QFTs are not integrable. It defines a continuous family of theories along a $T\bar T$ flow trajectory in the integrable quantum field theories space. Furthermore, the eigenvalue of energy and momentum of deformed theory are simply related to the expectation values of stress tensor, and the spectrum of the theories can be exactly obtained by solving a differential equation~\cite{Zamolodchikov2004}. Flow equation induced by $T\bar T$ deformation allows us to get the closed form formula for deformed Lagrangian, which relate to Nambu-Goto Lagrangian~\cite{Cavaglia2016, Bonelli:2018kik}. The deformation in general dimensions are also considered~\cite{Hartman:2018tkw, Taylor:2018xcy}. Irrelevant deformations, by definition, do not affect the IR behavior but will generally affect UV physics. Conformal field theory, as a UV complete framework, it attracts a big interesting to explore the deformed CFT, and $T\bar T$ deformed CFT might provide a route to extend the holographic duality not only in the frame of CFT. Some properties of $T\bar T$ deformed CFT are also studied. Based on the spectrum of deformed CFT, the partition function on the torus was obtained and modular properties have been discussed~\cite{Datta:2018thy, Aharony:2018bad}. 
\par
It is shown that the conserved charges remain conserved along the $T\bar T$ flow~\cite{Smirnov2017}. In a sense, the integrable deformation implies that it preserves an infinite amount of conserved charges, especially for the CFT, which have a conformal symmetry. As is known, the $T\bar T$ deformation breaks the conformal symmetry because the operator $T\bar T$ has conformal dimension $4$. However, it is reasonable to believe there still exist some other symmetries go beyond conformal symmetry for an integrable QFTs. From the renormalization group viewpoint, the $T\bar T$ flow defines a family of theories, and conformal field theory constitutes the fixed point of RG flow. One can guess the symmetry also changes following the $T\bar T$ flow. In case of CFT, the conformal symmetry leads to the traceless of stress tensor. It is turned out the deformation transforms the traceless into a trace relation, which depends on the deformation parameter $\mu$. Therefore, the trace relation may imply a symmetry of deformed CFT. We will give a detail illustration in section~\ref{sec: Sym}, especially for the background also depends on $\mu$. So the $T\bar T$ flow has an effect of changing the background of the boundary field theory, which is associated with different radial cutoff. This feature has some omens to associate $T\bar T$ deformed CFT to holographic renormalization group.         
\par
More interesting about the $T\bar T$ deformed CFT is the holographic aspects. McGough, Mezei and Verlinde proposed a novel idea, when study the $T\bar T$ deformed CFT~\cite{McGough2018}. It related the $T\bar T$ deformed CFT to AdS gravity, explicitly, the deformation corresponds to putting Dirichlet boundary condition in the bulk at finite radius. The deformation parameter is identified with inverse of cutoff radius square $\mu=\frac{24\pi}{c}\frac{1}{r^2_c}$. This means the deformation of CFT is equivalent to a radial cutoff in bulk, actually a UV/IR relation. Some holographic dictionaries have been established to support this relation, like the energy spectrum of deformed CFT is equal to the Brown-York's quasilocal energy and the RG equation is identical to Wheeler-DeWitt equation in 3D AdS gravity. It is a significative and challenging work to explore more holographic aspects of $T\bar T$ deformed CFT. Entanglement entropy~\cite{Chen:2018eqk, Murdia:2019fax, Jeong:2019ylz, Grieninger:2019zts, He:2019vzf, Donnelly:2019pie} and complexity~\cite{Chen:2019mis} have been studied. Path integral optimization about $T\bar T$ deformation and its geometric duality was explored~\cite{Jafari:2019qns}. In AdS$_3$/CFT$_2$, the symmetry plays an essential role, and asymptotic symmetry of AdS$_3$ gives  the Virasoro algebra~\cite{Brown1986}. However, we know little about the symmetry of $T\bar T$ on both boundary field theory and cutoff gravity side. We mainly focus on the symmetry of the $T\bar T$ deformed CFT and give a holographic interpretation in this paper.    
\par
The cutoff AdS$_3$ with Dirichlet boundary condition means boundary metric depends on cutoff radius. According to this proposal, the boundary metric is not a fixed one, but runs with the cutoff radius. This inspires us to consider the $T\bar T$ deformed CFT living on a dynamical background. But this feature is unintelligible from field theory side. In fact, Cardy has discovered the $T\bar T$ deformation is equivalent to the random metric~\cite{Cardy2018}, which can give a description of $T\bar T$ deformed CFT with a dynamical background. The flow equation of metric can also be obtained at the saddle point. In this paper, we consider a probable symmetry of $T\bar T$ deformed CFT, which preserve the trace relation of $T\bar T$ deformed CFT. The deformed conformal killing equation is obtained. According to the flow equation of metric, we find the deformation effect can be expressed by adding a change rate of metric along parameter $\mu$ term on the conformal killing equation, and conformal symmetry coincides with the fixed point. On the gravity side, we restudy the Dirichlet boundary condition and find it gives a new asymptotic symmetry of AdS$_3$, which corresponds to the $T\bar T$ symmetry. Moreover, we derive that the stress tensor equals to Brown-York's quasilocal stress tensor. As a specific realization, we consider a simple nontrivial case, the BTZ black hole. We get the $T\bar T$ deformed conformal killing vectors and the conserved charges are also calculated. However, we can only get two independent conserved charges.
\par
This paper is organized as follows: We restudy the $T\bar T$ deformed CFT on dynamical background, which runs with deformation parameter in section~\ref{sec: TT in dynamical background}. In section~\ref{sec: Sym}, we propose a symmetry of $T\bar T$ deformed CFT based on the trace relation. The symmetry can be described by a deformed conformal killing equation. We give a holographic interpretation of the deformed conformal killing equation on the gravity side in  section~\ref{sec: Hol}, which coinsides with a boundary condition of AdS$_3$. The holographic stress tensor is also considered. In section~\ref{sec: BTZ}, a specific example, BTZ black hole, is studied. Based on this equation, we get the $T\bar T$ deformed conformal killing vector and conserved charges are also considered. Conclusion, discussion about symmetry algebra and Chern-Simons formalism are given in section~\ref{sec: con}.      
\section{$T\bar T$ deformed CFT with dynamical background}
\label{sec: TT in dynamical background}
In this section, we review some main results about $T\bar T$ deformed CFT and its holographic aspects, which would be useful for discussing the symmetry in this paper. Especially, the background metric runs with the deformation parameter $\mu$, instead of a static one. We start from the definition of $T\bar T$ deformed theory living on an arbitrary background.      
\par 
The action of a $T\bar T$ deformed QFT is defined by the $T\bar T$ flow 
\begin{align}
\label{S_flow}
\frac{\partial S}{\partial \mu}=\frac{1}{2}\int d^2x(\sqrt{g} T\bar T)_{\mu}, 
\end{align}
where the operator $T\bar T\equiv(T^{ij}T_{ij}-T^2)=-\frac{1}{2}\varepsilon_{ik}\varepsilon_{jl}T^{ij}T^{kl}$. The label $\mu$ denotes a quantity depends on the deformation parameter $\mu$, especially the background metric $(g_{ij})_{\mu}$. Usually, one may consider the theory living on a Minkowski spacetime or a static curved background. However, the metric is not fixed one in our configuration, but depends on the deformation parameter. Dimensional analysis yields the parameter $\mu$ has dimension of $[M]^{-2}$. Besides, in general, for a theory with a single mass scale $\lambda$, its action has the relation 
\begin{align}
\lambda \frac{\partial S}{\partial \lambda}=\frac{1}{2}\int d^2x\sqrt{g} T^i_i. 
\end{align}
By setting $\mu=1/\lambda^2$, one can get an important trace relation of $T\bar T$ deformed CFT
\begin{align}
\label{trace relation}
T^i_i=-2\mu T\bar T,
\end{align}     
which would imply the symmetry of $T\bar T$ deformed CFT we will discuss in section 3. For $\mu=0$, the trace relation reduces to traceless of stress tensor, and this is consistent with CFT case. So the trace relation is just the generalization of traceless of a CFT along the $T\bar T$ flow.
\par
For an infinitesimal variation of $\mu$, the change in action is 
\begin{align}
S(\mu+\delta \mu)-S(\mu)=\frac{\delta\mu}{2}\int\sqrt{g}\varepsilon_{ik}\varepsilon_{jl}T^{ij}T^{kl}.
\end{align}
This will lead to a variation of path integral by inserting a factor $e^{-\delta S}$. It can be expressed as a functional integral through Hubbard-Stratonovich transformation
\begin{align}
&e^{-\frac{\delta\mu}{2}\int\sqrt{g}\varepsilon_{ik}\varepsilon_{jl}T^{ij}T^{kl}}\propto  \int [dh]e^{\frac{1}{2\delta\mu}\int\sqrt{g}\varepsilon^{ik}\varepsilon^{jl}h_{ij}h_{kl}d^2x-\int \sqrt{g}h_{ij}T^{ij}d^2x}.
\end{align}
The value of functional integral can be obtained at saddle point approximate
\begin{align}
&T_{ij}=\frac{1}{2\delta\mu}\varepsilon_{ik}\varepsilon_{jl}h^{ij}=\frac{1}{2\delta\mu}(g_{ij}h-h_{ij}).
\end{align}
By the definition of $T_{ij}$, $h_{ij}$ is an infinitesimal change of the metric. Then above equation can be expressed as 
\begin{align}
\label{g_variation}
h_{ij}=\delta g_{ij}=2\delta\mu(g_{ij}T-T_{ij}).
\end{align}
At last, we get the flow equation about background metric 
\begin{align}
\label{g_flow}
\frac{\partial}{\partial{\mu}}g_{ij}&=-2(T_{ij}-Tg_{ij}).
\end{align}
This means that the background metric also runs with the parameter $\mu$. If the initial theory, i.e. $\mu=0$, is a 2D CFT, this result holds for a small $\mu$ or large $c$ limit, which means $c$ is large enough but fix the $\mu c$. This conclusion was obtained by Cardy via the viewpoint of random geometry~\cite{Cardy2018}, and the correlation function has been calculated~\cite{Aharony2018,He:2019vzf}. The flow equation of metric can also be obtained by the variational principle. In brief, the variation of action can be written as
\begin{align}
\label{S_variation}
\delta S=-\frac{1}{2}\int d^2x\sqrt{g}T^{ij}\delta g_{ij}.
\end{align} 
Combining \eqref{S_flow}, \eqref{S_variation} and $\delta\partial_{\mu}S=\partial_{\mu}\delta S$, one can also get the flow equation. This variational principle approach was proposed by Guica and Monten in~\cite{Guica2019}. In addition, through resolving the flow equations, the stress tensor of $T\bar T$ deformed CFT can be obtained, which coincide with the Brown-York's quasilocal stress tensor~\cite{Brown1993}.
\par
The deformed Lagrangian in closed form has been obtained in~\cite{Bonelli:2018kik} by reformulating the $T\bar T$ deformation as a functional equation. We can not get a closed form of the deformed Lagrangian with a unfixed metric, because the $T\bar T$ flow can just determine the action. If the metric also varies with $\mu$, which means there is a varied integral measure. So we can not get a deformed Lagrangian formula at present.
\par
Something more interesting is the holographic aspects of $T\bar T$ deformed CFT. There is a proposal that the $T\bar  T$ deformed CFT is dual to a cutoff AdS at $r=r_c$ with Dirichlet boundary condition~\cite{McGough2018,Kraus2018}. The bulk cutoff $r=r_c$ relates to the deformation parameter by
\begin{align}
\label{GMV relation}
\mu=\frac{24\pi}{c}\frac{1}{r_c^2}.
\end{align}
This means different cutoff in bulk is equivalent to different deformation parameter on the boundary field. But in this paper, we would not fix the parameter $\mu$, so we may write $\mu=\frac{24\pi}{c}\frac{1}{r^2}$. All the conclusions we get still hold when setting $r=r_c$. An important aspect is that the induced metric on the finite radial boundary also depends on $r$ (or $\mu$). This feature matches the random geometry we reviewed above. Therefore, in order to explore the holographic aspects, it is essential to consider the dynamical background. 
\par
We list some results here, which would be instructive for us when discuss the conserved charge in section~\ref{sec: BTZ}. For detail please see~\cite{McGough2018,Kraus2018}. The spectrum of $T\bar T$ deformed CFT on a cylinder with circumference $L$ can be obtained by solving a differential equation, which takes the form
\begin{align}
\label{TT spectrum}
E_n\equiv L\langle n|T_{tt}|n\rangle=\frac{2L}{\mu}\left(1-\sqrt{1-\frac{2\pi\mu}{L^2}M_n+\frac{\mu^2J_n^2}{L^4}}\right),
\end{align} 
where $M_n=\Delta_n+\bar\Delta_n-c/12$ and $J_n=\Delta_n-\bar\Delta_n$. On the gravity side, the quasilocal proper energy of BTZ black hole, defined by Brown and York~\cite{Brown1993, Brown1994}, is
\begin{align}
\label{BY energy}
E\equiv\int\sqrt{g_{\phi\phi}}\tau_{ij}u^iu^jd\phi=\frac{r}{4G}\left(1-\sqrt{1-\frac{8G M}{r^2}+\frac{16G^2J^2}{r^4}}\right),
\end{align}
where $u^{i}$ is the unit vector normal to a time slice and $\tau_{ij}$ is Brown-York stress tensor of gravity. The proper size of the spatial circle on the boundary is $2\pi r$, and the circumference of cylinder usually set to $L=2\pi$. Moreover, the quasilocal energy tends to $M$, when $r\to \infty$. Noting the relation~\eqref{GMV relation}, one can identify
\begin{align}
LE_n=2\pi rE.
\end{align}
What makes more sense is that we can write it as
\begin{align}
\label{spectrum grav}
E_{n}=r\int\sqrt{g_{\phi\phi}}\tau_{ij}u^iu^jd\phi.
\end{align} 
The angular momentum $J_{n}$ can also be obtained from gravity side by the definition of quasilocal conserved charge 
\begin{align}
\label{angular grav}
J_n=J\equiv\int\sqrt{g_{\phi\phi}}\tau_{ij}u^i\xi^jd\phi,
\end{align} 
where the $\xi^{i}$ is the killing vector associate with $J$.  
\section{Symmetry of $T\bar T$ deformed CFT}
\label{sec: Sym}
In classical field theory, the invariance of the action under an infinitesimal transformation implies a symmetry. For a conformal invariant theory, things are more intriguing. By definition, the conformal transformation leaves the metric invariant up to a scaling factor 
\begin{align}
\label{conformal variation}
\delta g_{ij}=\omega(x)g_{ij}. 
\end{align}
Then, the variation of action can be written as     
\begin{align}
\label{SCFT_variation}
&\delta S_{CFT}=-\frac{1}{2}\int d^2x\sqrt{g} T^{ij}\delta g_{ij}=-\frac{1}{2}\int d^2x\sqrt{g} T^{i}_{i}\omega(x).
\end{align}
Thus the conformal invariance leads to the traceless of stress tensor $T^i_i=0$. Conversely, the traceless of stress tenser implies conformal symmetry of a quantum field theory. As for $T\bar T$ deformed CFT, the conformal symmetry is abandoned. However, it has another property analogous to the traceless, namely the trace relation. From the viewpoint of flow equation, the deformed CFT is defined by $T\bar T$ flow, and the $T\bar T$ flow transforms the traceless into the trace relation along the parameter $\mu$. In the case of CFT, that is $\mu=0$, the trace relation reduced to traceless of stress tensor, which implies the conformal symmetry. Therefore, we may conjecture the trace relation would imply the symmetry of $T\bar T$ deformed CFT. 
\par
To obtain the symmetry from trace relation, we can rewrite it as
\begin{align}
T^i_i+2\mu T\bar T=T^{ij}[g_{ij}+2\mu(T_{ij}-g_{ij}T^{k}_{k})]=0.
\end{align}
Analogy to the conformal case, we replace the traceless of stress tensor by the trace relation in variation of action, it ends up with  
\begin{align}
\label{action principle}
\delta S(\mu)&=-\frac{1}{2}\int d^2x\sqrt{g}T^{ij}\delta g_{ij}\nonumber\\
&=-\frac{1}{2}\int d^2x\sqrt{g}T^{ij}[g_{ij}+2\mu(T_{ij}-g_{ij}T^{k}_{k})]\omega(x)=0.
\end{align}
With this ansatz, the trace relation leads to invariance of action, namely a symmetry. That is to say the variation of metric, under this symmetry transformation, would be 
\begin{align}
\label{metric variation 1}
\delta g_{ij}=[g_{ij}+2\mu(T_{ij}-g_{ij}T^{k}_{k})]\omega(x),
\end{align}
where the $\omega(x)$ is the Weyl-like factor, an arbitrary function of $x$. The second term on the right hand side comes from the $T\bar T$ deformation. However, it is not clearly to understand what is the effect of $T\bar T$ deformation on geometry from above equation, because the deformation depends not only metric but also stress tensor. Fortunately, the second term can be written in another form as we have mentioned in section 2. The basic idea is the background is not fixed, but runs with deformation parameter $\mu$. The deformation term is just the variation of metric along parameter $\mu$, according to \eqref{g_flow}. Therefore, the variation of metric in $T\bar T$ deformation becomes     
\begin{align}
\label{metric variation 2}
\delta g_{ij}=(g_{ij}-\mu\frac{\partial g_{ij}}{\partial \mu})\omega(x),
\end{align}
from which, we can easily see that the deformation lead to an additional term $\mu\partial_{\mu}g_{ij}$ compared with conformal killing equation~\eqref{conformal variation}. This term is just the rate of change of the metric along the parameter $\mu$, and the fixed point consistent with the conformal symmetry.
\par
We have to point out the difference between \eqref{metric variation 2} and \eqref{g_variation}. In \eqref{g_variation}, the variation of metric with respect to $\mu$, which yields the flow equation of metric \eqref{g_flow}. This result was also got via the variational principle in~\cite{Guica2019}. By solving the flow equation, the authors get the general solution of metric that is consistent with Fefferman-Graham gauge. Further, this solution implies a mixed non-linear boundary condition for the asymptotic metric, which leads to two commuting copies of a state-dependent Virasoro algebra~\cite{Guica2019}. While the variation of metric in  \eqref{metric variation 2} implies not only a flow equation of metric, but also a symmetry for $T\bar T$ deformed CFT. Because \eqref{metric variation 2} comes from the invariance of action \eqref{action principle}, which means the action arrives extremum. In fact, \eqref{metric variation 2} is a generalization of conformal killing equation, and this is one of our main results. A holographic interpretation of this equation and a solution with regard to BTZ black hole would be given in later part of this paper.
\par 
Besides, one should note that we do not write the variation of metric in Lie derivative form, because it depends on stress tensor which may refer to the non-geometry quantities. Generally, this is a gauge transformation, which may not be realized by an infinitesimal diffeomorphism $x^i\to x^i+\xi^{i}$. However, we would see that there actually exist coordinates transformation on the boundary from the bulk asymptotic symmetry to achieve the variation of metric. So one can replace $\delta g_{ij}$ by $\mathcal L_{\xi}g_{ij}$, and above equation becomes a $T\bar T$ deformed conformal killing equation. We will give a detail illustration in next section. In addition, this $T\bar T$ deformed conformal killing equation is not covariant under a coordinate transformation refers to $\mu$ because of the second term. In fact, for two dimensional case, one can always transform the background to Minkowski spacetime in new coordinates. But the equation \eqref{metric variation 2} should also change. So it does not simplify the problem, when one tries to solve this equation.  
\par
This deformed conformal killing equation is obtained based on the trace relation. However, this trace relation just holds on classical level. The quantum effect may be lead to an anomaly term like \eqref{TT trace anomaly}. As the deformed killing equation is not covariant, this equation \eqref{metric variation 2} highly depends on a special coordinate. On gravity side, we will show this special coordinate coincides with Fefferman-Graham gauge of AdS$_3$.
\par
With this deformed killing equation, we can work out the Ward identity. Given an infinitesimal coordinates transformation, we have 
\begin{align}
\label{deformed_omega}
\nabla_{i}\xi_j+\nabla_{j}\xi_i=(g_{ij}-\mu\frac{\partial g_{ij}}{\partial \mu})\omega(x).
\end{align}
Then the Ward identity can be expressed as
\begin{align}
\label{ward identity}
\delta_{\xi}\langle X\rangle=\xi_{i}\nabla_{j}\langle T^{ij}X\rangle+\frac{ \nabla_i\xi^i}{2-\mu\partial_{\mu}\ln g}\langle(T^i_i-\mu\partial_{\mu}g_{ij}T^{ij}) X\rangle,
\end{align}
where $X$ stands for a string of fields in deformed theory, $g$ represents determinant of metric. For  $\mu=0$, it gives the conformal Ward identity. However, the left hand side of this equation still unknown in deformed CFT, because it is not clear how the fields vary under coordinates transformation. For further study on this Ward identity, a representation theory for the deformed symmetry group is needed.
\section{Holographic interpretation}
\label{sec: Hol}
In this section, we would explore holographic aspects of the symmetry of $T\bar T$ deformed CFT. We will show that the $T\bar T$ symmetry corresponds to a new asymptotic boundary condition of AdS$_3$. Moreover, the stress tensor of boundary field theory can also be obtained from the gravity side.
\subsection{New boundary condition of AdS$_3$}
It was proposed that the $T\bar T$ deformed 2D CFT corresponds to a cutoff AdS$_3$. So the symmetry of the $T\bar T$ should have a holographic interpretation on the gravity side, the $T\bar T$ deformed conformal killing equation in particular. We begin with the most general solution of Einstein equation with a negative cosmological constant, which can be expressed in Fefferman-Graham expansion~\cite{SKENDERIS2000316}
\begin{align}
ds^2&=G_{\mu\nu}dx^{\mu}dx^{\nu}=\frac{l^2}{r^2}dr^2+\gamma_{ij}dx^idx^j, \nonumber\\
\label{induced metric F-G}
\gamma_{ij}&=r^2(g^{(0)}_{ij}(x)+\frac{1}{r^2}g^{(2)}_{ij}(x)+\frac{1}{r^4}g^{(4)}_{ij}(x)).
\end{align}
The Einstein equation also gives a constraint
\begin{align}
g^{(4)}_{ij}=\frac{1}{4}g^{(2)}_{ik}g^{(0)kl}g^{(2)}_{jl}.
\end{align}
From holographic correspondence, the boundary metric $\gamma_{ij}$ is not well defined at spatial infinity $r\to\infty$, but one can choose a regularization factor $1/r^2$ to cancel the singularity~\cite{Witten:1998qj}, such that the boundary metric is well defined, that is $g^{(0)}_{ij}$ for a Poincar\'e patch. Generally, one can choose $g^{(0)}_{ij}=\eta_{ij}$, then there will be a conformal field theory on the boundary at spatial infinity. Now we suppose the boundary field theory lives on the regularized background metric
\begin{align}
\label{boundary metric}
g_{ij}=\frac{\gamma_{ij}}{r^2}=g^{(0)}_{ij}(x)+\frac{1}{r^2}g^{(2)}_{ij}(x)+\frac{1}{r^4}g^{(4)}_{ij}(x),
\end{align}  
which is also well defined even at spatial infinity. 
\par
In the context of AdS$_3$/CFT$_2$, Brown and Henneaux found an asymptotic boundary condition of AdS$_3$, and the asymptotic symmetry algebra is consistent with Virasoro algebra with a nontrivial central charge~\cite{Brown1986}. So the conformal symmetry of boundary field theory can be described by a asymptotic symmetry in bulk. Some other asymptotic boundary conditions at spatial infinity have been also explored, like BMS$_3$ and $U(1)$ Kac-Moody-Virasoro algebra~\cite{Guica:2008mu, Barnich:2010eb, Compere:2013bya, Grumiller:2016pqb}. But we now consider the theory at finite boundary, a $T\bar T$ deformed CFT, which is dual to gravitational theory restricted on a compact sub-region of AdS spacetime. Brown and Henneaux's boundary condition just fix the leading order $g^{(0)}_{ij}$. In~\cite{Guica2019}, Guica and Monten have proposed a mixed boundary condition for the $T\bar T$ deformation, namely setting the initial metric as $g^{(0)}_{ij}$ and adding higher order term of parameter $\mu$ at cutoff surface for the deformation, that is actually a Fefferman-Graham expansion. This mixed boundary condition exactly corresponds to fixing the induced metric on a constant radius surface. Nevertheless, we mainly focus on the deformed conformal killing equation in this paper.
\par
In our configuration, we would like to consider the initial background metric takes the form \eqref{induced metric F-G}, instead of $g^{(0)}_{ij}$ in~\cite{Guica2019}. On the cutoff boundary, we propose a boundary condition that the variation of metric was allowed as
\begin{align}
\label{boundary condition}
\delta G_{rr}=0,\quad \delta G_{ri}=O(\frac{1}{r}),\quad \delta G_{ij}=\delta\gamma_{ij}=0.
\end{align}  
This boundary condition is different from the mixed boundary condition in ~\cite{Guica2019}, because the variation of metric is not just a flow equation but also a special transformation. A specific coordinate transformation is discussed in section 5 for BTZ black. We now show that the symmetry of $T\bar T$ deformed CFT, particularly the equation \eqref{metric variation 2}, corresponds to this boundary condition for AdS$_3$ exactly. Consider an infinitesimal diffeomorphism transformation in bulk $\zeta^{\mu}=(\zeta^{r},\xi^{i})$, where the $\xi^{i}$ represents a diffeomorphism on the boundary at constant $r$. Under this infinitesimal transformation, the variation of metric could be expressed by Lie derivative. For the $rr$ component
\begin{align}
\delta G_{rr}=\mathcal L_{\zeta}G_{rr}=\frac{l}{r}\partial_{r}\xi^{r}=0,
\end{align}
where we have set $\xi^{r}=2l\zeta^{r}/r$. The solution of this equation is $\xi^{r}=\xi^{r}(x)$, This means $\xi^{r}$ is an arbitrary function which does not depend on the radial coordinate $r$. Then we turn to consider the variation of induced metric on the boundary
\begin{align}
\delta \gamma_{ij}=\mathcal L_{\zeta}\gamma_{ij}&=\zeta^{\mu}\partial_{\mu}\gamma_{ij}+\gamma_{ik}\partial_{j}\zeta^{k}+\gamma_{kj}\partial_{i}\zeta^{k}\nonumber\\
&=\xi^{r}\frac{r}{2l}\partial_{r}\gamma_{ij}+\xi^{k}\partial_{k}\gamma_{ij}+\gamma_{ik}\partial_{j}\xi^{k}+\gamma_{kj}\partial_{i}\xi^{k}\nonumber\\
&=-\xi^{r}K_{ij}+\mathcal L_{\xi}\gamma_{ij}.
\end{align}
Here the definition of extrinsic curvature is used   
\begin{align}
K_{ij}=-\frac{1}{2}\mathcal L_{n}\gamma_{ij}=-\frac{r}{2l}\partial_{r}\gamma_{ij},
\end{align}
where $n=(r/l,0,0)$ is the  normal vector to surfaces of constant $r$. Therefore, the boundary condition $\delta \gamma_{ij}=0$ would yield the equation
\begin{align}
\label{induced metric variation}
\mathcal L_{\xi}\gamma_{ij}&=\xi^{r}K_{ij}.
\end{align}
This equation shows the variation of induced metric equals the extrinsic curvature up to a scaling factor $\xi^{r}$, which just depends on the boundary coordinates $x^{i}$. The $ri$ components $\delta G_{ri}=O(1/r)$ illustrate an asymptotic behavior, and it might give boundary dynamics. We would not discuss this problem in this paper. When $r\to \infty$, the metric \eqref{induced metric F-G} becomes Poincar\'e patch and the symmetry is a isometry of AdS$_3$. Therefore, the asymptotic killing equation \eqref{induced metric variation} is obtained from the boundary condition. On the other hand, $T\bar T$ deformed conformal killing equation is also got in last section. We, then, will show that this asymptotic symmetry is just the symmetry of $T\bar T$ deformed CFT living on a surface of constant $r$ with metric $g_{ij}$.
\par 
In fact, we should note that the boundary metric is regularized induced metric by $g_{ij}=\gamma_{ij}/r^2$, and that is the essential difference between \eqref{metric variation 2} and \eqref{induced metric variation}. The \eqref{induced metric variation} can be written in terms of $g_{ij}$ 
\begin{align}
\mathcal L_{\xi}\gamma_{ij}&=\xi^{r}K_{ij} =-\xi^{r}\frac{r}{2l}\partial_{r}(r^2g_{ij})\nonumber\\
&=-\xi^{r}\frac{r}{2l}(2rg_{ij}+r^2\partial_{r}g_{ij})\nonumber\\
&=-\xi^{r}\frac{r^2}{l}(g_{ij}-\mu\frac{\partial g_{ij}}{\partial \mu}), 
\end{align}
where the relation $\mu\sim\frac{1}{r^2}$ is used in above steps, and the coefficient is dispensable. After dividing out $r^2$ on both sides, it ends up with 
\begin{align}
\label{TT killing equation}
\mathcal L_{\xi}g_{ij}&=-\xi^{r}\frac{1}{l}(g_{ij}-\mu\frac{\partial g_{ij}}{\partial \mu}),
\end{align}
which is same as \eqref{metric variation 2} with Weyl-like factor $\omega(x)=-\xi^{r}(x)/l$. The nontrivial term $\mu\partial_{\mu}g_{ij}$ comes from the $T\bar T$ deformation. It has a contribution to this deformed conformal killing equation, if we consider the high order terms of the boundary metric, because the high order term depends on $\mu$ (or $r$). The leading term $g^{(0)}_{ij}$ does not depend on $\mu$, which implies the conformal symmetry. When we push the boundary to spatial infinity, the high order terms tend to vanish, and the symmetry reduces to conformal symmetry. 
\par
The deformed conformal killing equation is obtained from the AdS$_3$ with a radial cutoff boundary. Like the boundary field perspective, this is also coordinates dependent. Therefore, our discussion is based on Fefferman-Graham gauge. However, the solution of AdS$_3$ terminates after fourth order in Fefferman-Graham expansion, thus we can get the symmetry transformation at finite radial  exactly. For a pure AdS$_3$, Poincar\'e patch in Fefferman-Graham gauge, the higher order terms vanish, namely  $g^{(2)}_{ij}=0$ and $g^{(4)}_{ij}=0$. The cutoff is a trivial case, because  the boundary metric always flat for any radial cutoff. Then, the deformed conformal killing equation always reduces to original one. In order to get a nontrivial case, we must consider the high order correction of Fefferman-Graham expansion. A simple nontrivial case is BTZ black hole background, we will study it in section~\ref{sec: BTZ}. 
\par
In terms of Fefferman-Graham expansion of boundary metric $g_{ij}$, according to the deformed conformal killing equation \eqref{boundary metric} and \eqref{TT killing equation}, one can get 
\begin{align}
\mathcal L_{\xi}\left(g^{(0)}_{ij}+\frac{1}{r^2}g^{(2)}_{ij}+\frac{1}{r^4}g^{(4)}_{ij}\right)=\left(g^{(0)}_{ij}-\frac{1}{r^4}g^{(4)}_{ij}\right)\omega(x),
\end{align}
from which we can see that the usual conformal symmetry is just the leading term of above equation, namely $\mathcal L_{\xi}g^{(0)}_{ij}=g^{(0)}_{ij}\omega(x)$. The Brown-Henneaux's boundary condition actually preserves the leading order of $O(1/r)$. The $T\bar T$ symmetry includes the high order correction, and the  correction terminates after fourth order in case of AdS$_3$. Therefore, this new boundary condition just replaces the asymptotic symmetry of $\gamma_{ij}$ and $G_{rr}$ by exactly killing equations, but keeps the asymptotic symmetry of $G_{ri}$. That is we have remarked in \eqref{boundary condition}.
\par
We finish this discussion about deformed conformal killing equation with a few comments about sign of deformation parameter. In our discussion on boundary field theory in section~\ref{sec: Sym}, the derivatation holds for both sign of deformation parameter. However, these may have different holographic physical interpretations. For positive sign $\mu>0$, it is known that the deformation can be interpreted as a cutoff AdS$_3$ with relation \eqref{GMV relation}~\cite{McGough2018}, this is the case we mainly focus on. As for $\mu<0$, it corresponds little string theory in the asymptotically linear dilaton spacetime~\cite{Giveon:2017nie}, which is not associated with our discussion, for a cutoff AdS$_3$, in this section. Therefore our discussion in the holographic part just for $\mu>0$, but the deformed conformal killing equation is valid for both signs of the deformation parameter.
\subsection{Holographic stress tensor}  
\label{subs: stress tensor}
As the $T\bar T$ deformed conformal killing equation relates to the stress tensor of boundary field theory and the symmetry also has a holographic description. So the stress tensor might be calculated from the gravity side.  In this subsection, we would derive a holographic duality of stress tensor form the deformed conformal killing equation, that is the stress tensor of $T\bar T$ deformed CFT equals to the Brown-York's quasilocal stress tensor on the cutoff surface, which have been got from the solution of flow equation in~\cite{Guica2019}. The trace anomaly of $T\bar T$ deformed CFT is also considered. The stress tensor duality would be used for discussing the quasilocal conserved charge of BTZ black hole in section~\ref{sec: BTZ}. 
\par
We begin with a $T\bar T$ deformed CFT placing at a finite radial surface distance $r$ from the center of the bulk. So the background metric depends on $\mu$. From \eqref{g_flow}, we can get the stress tensor relation of $T\bar T$ deformed CFT
\begin{align}
\label{boundary stress tensor relation}
\hat T_{ij}\equiv T_{ij}-T g_{ij}=-\frac{1}{2}\partial_{\mu}g_{ij}.
\end{align} 
The $T_{ij}$ represents the stress tensor of boundary field theory and we will use $\tau_{ij}$ to denote a holographic stress tensor from gravity side. It is a little complicated to express out the stress tensor $T_{ij}$ in terms of metric. We here would calculate the similar quantity $\hat \tau_{ij}\equiv\tau_{ij}-\tau\gamma_{ij}$ on the gravity side. Actually, the Brown-York's quasilocal stress tensor of gravity is defined as~\cite{Brown1993, Brown1994} 
\begin{align}
\label{BY stress tensor}
\tau_{ij}=\frac{2}{\sqrt{\gamma}}\frac{\delta S_{grav}}{\delta\gamma^{ij}}=-\frac{1}{8\pi G}(K_{ij}-K\gamma_{ij}-\frac{1}{l}\gamma_{ij}),
\end{align} 
where we have added  the last term to cancel divergences on the spatial infinity boundary. Taking trace and substituting back the equation, we arrive at
\begin{align}
\label{bulk stress tensor relation 1}
\hat\tau_{ij}\equiv\tau_{ij}-\tau\gamma_{ij}=-\frac{1}{8\pi G}(K_{ij}+\frac{1}{l}\gamma_{ij}).
\end{align} 
There is still something different from \eqref{boundary stress tensor relation}. However, noting that the extrinsic curvature can be written in terms of $g_{ij}$ 
\begin{align}
K_{ij}=-\frac{r}{2l}\partial_{r}\gamma_{ij}=-\frac{1}{l}\gamma_{ij}-\frac{r^3}{2l}\partial_{r}g_{ij},
\end{align}
plugging into \eqref{bulk stress tensor relation 1}, one can get 
\begin{align}
\label{bulk stress tensor relation 2}
\hat\tau_{ij}=\frac{r^3}{16\pi Gl}\partial_{r}g_{ij}.
\end{align} 
If we identify $\mu\sim1/r^2$, explicitly $\mu=4\pi Gl/r^2$, the stress tensors on both sides are equivalent  
\begin{align}
\label{stress tensor cor}
\hat T_{ij}=\hat\tau_{ij}\Leftrightarrow T_{ij}=\tau_{ij}.
\end{align} 
Therefore we derive a holographic dictionary about stress tensor. Hence, by applying the Brown-Henneaux relation $c=3l/2G$~\cite{Brown1986}, we obtain 
\begin{align}
\label{GMV relation 2}
\mu=\frac{6\pi l^2}{c}\frac{1}{r^2},
\end{align} 
which is just the relation \eqref{GMV relation}, up to an unessential constant, because we use the different convention of the metric form. In fact, the same quantity is marked as $\tilde\mu$ in~\cite{McGough2018}.
\par
If we consider the Fefferman-Graham expansion of the induced metric \eqref{induced metric F-G}, the stress tensor can be calculated from gravity side
\begin{align}
\label{holographic stress tensor}
\hat T_{ij}=\hat \tau_{ij}=-\frac{1}{8\pi Gl}\left(g^{(2)}_{ij}+\frac{2}{r^2}g^{(4)}_{ij}\right).
\end{align} 
According to the holographic dictionary of AdS$_3$/CFT$_2$~\cite{Balasubramanian1999, deHaro2001}, the leading order coincide with the CFT case. When we push the boundary to spatial infinity, the metric reduces to $g^{(0)}_{ij}$. Then the trace of stress tensor is obtained  by contracting with $g^{(0)ij}$
\begin{align}
\label{stress tensor trace}
T^i_i=\frac{1}{r^2}\tau^i_i=\frac{1}{8\pi Gl}\text{Tr}[(g^{(0)})^{-1}g^{(2)}].
\end{align}   
Einstein equation perturbatively gives~\cite{Henningson:1998gx}
\begin{align}
\text{Tr}[(g^{(0)})^{-1}g^{(2)}]=\frac{l^2}{2}R,
\end{align}
combining with the Brown-Henneaux relation, \eqref{stress tensor trace} ends up with the well-known trace anomaly of CFT
\begin{align}
\label{trace anomaly}
T^i_i=\frac{c}{24\pi}R.
\end{align}
As for the higher term, we should contract \eqref{holographic stress tensor} with $g^{ij}$ instead of $g^{(0)ij}$. It is a complicated perturbation calculation, because the scalar curvature also changes with parameter $\mu$ or for different cutoff $r$. However, with the correspondence of stress tensor, we can get the trace anomaly exactly from gravity side. 
\par
In fact, the $T\bar T$ operator can be calculated from gravity side
\begin{align}
T\bar T=\tau\bar\tau=\hat\tau_{ij}\tau^{ij}=\frac{1}{(8\pi G)^2}\left(K^{ij}K_{ij}-K^2-\frac{2}{l}K-\frac{2}{l^2}\right).
\end{align}
Noting the Gauss-Codazzi relation, we obtain 
\begin{align}
T\bar T=-\frac{1}{(8\pi G)^2}\left(R+\frac{2}{l}K+\frac{4}{l^4}\right).
\end{align}
Moreover, the trace of stress tensor is 
\begin{align}
T^i_i=\frac{1}{r^2}\tau^i_i=-\frac{1}{8\pi Gr^2}\left(K+\frac{2}{l}\right).
\end{align}
Combining these equations, we arrive at the relation
\begin{align}
\label{TT trace anomaly}
T^i_i=\frac{c}{24\pi}R-2\mu T\bar T,
\end{align}
The trace anomaly of $T\bar T$ deformed CFT, which is also obtained in~\cite{Guica2019, Shyam:2017znq} from the field theory side. According to the stress tensor correspondence, we can reproduce it from gravity side.
\par 
The trace anomaly, in a sense, is the exact RG equation. It corresponds to Wheeler-DeWitt equation in 3D AdS gravity. Actually, the Brown-York's stress tensor is the canonical momentum in ADM formalism, up to a counterterm. In classical level, if one replaces the boundary stress tensor by Brown-York's stress tensor, the trace relation exactly gives the Hamiltonian constraint. After quantization, one can get that holographic RG equation which is equivalent to Wheeler-DeWitt equation~\cite{McGough2018, Donnelly:2019pie}. It has not been clear why this holographic cutoff exactly looks the same from the QFT perspective. But this is an evidence that this holographic cutoff of AdS is precise described by the $T\bar T$ deformed CFT on the boundary. The boundary metric is exactly induced metric of AdS$_3$ solution because of Hamiltonian constraints, namely a Dirichlet boundary condition. 
\section{BTZ black hole}
\label{sec: BTZ}
The BTZ black hole~\cite{Banados:1992wn, Banados:1992gq} is a good realization of cutoff AdS$_3$ and $T\bar T$ deformed 2D CFT correspondence. The spectrum  can be obtained by considering the quasilocal energy defined on a surface at finite radial location. In this section, we would explore more about $T\bar T$ deformed CFT from BTZ black hole, including $T\bar T$ symmetry and conserved charges. 
\par
There is a useful coordinate system in Fefferman-Graham gauge for BTZ black hole~\cite{banados1999three} 
\begin{align}
\label{btz metric}
ds^2=&\frac{l^2}{r^2}dr^2+r^2dzd\bar z+\frac{L\bar L}{r^2}dzd\bar z+Ldz^2+\bar Ld\bar z^2,
\end{align}
where $z=t-i\phi,\bar z=t+i\phi$ and $r$ is a Rindler-like proper radial coordinate, $L$ and $\bar L$ are constants related to the parameters of BTZ black hole $M$ and $J$ by
\begin{align}
L=&\frac{(r_++r_-)^2}{2l}=\frac{1}{2}(M+J),\bar L=\frac{(r_+-r_-)^2}{2l}=\frac{1}{2}(M-J).
\end{align}
In this coordinate system, the induced metric on a surface at finite radial can be expressed as  
\begin{align}
\label{induced metric of BTZ}
\gamma_{ij}=(r^2+\frac{L\bar L}{r^2})\bar\eta_{ij}+\bar\delta_{ij},
\end{align}
where we have defined 
\begin{align}
\bar\eta_{ij}=\left(\begin{matrix}
	0 & 1/2\\
	1/2 & 0\\
\end{matrix}\right),
\bar\delta_{ij}=\left(\begin{matrix}
	L & 0\\
	0 & \bar L\\
\end{matrix}\right).
\end{align}
The extrinsic curvature of the boundary can be calculated   
\begin{align}
K_{ij}=\frac{1}{l}(r^2-\frac{L\bar L}{r^2})\bar\eta_{ij},
\end{align}
then the $T\bar T$ deformed conformal killing equation \eqref{induced metric variation} becomes 
\begin{align}
\mathcal L_{\xi}\gamma_{ij}&=\partial_i\xi_{j}+\partial_{j}\xi_{i}\nonumber\\
&=\xi^{r}(x)\left(r^2-\frac{L\bar L}{r^2}\right)\bar\eta_{ij}.
\end{align}
We use partial derivative here, because the Levi-Civita connection of $\gamma_{ij}$ vanishes. The solutions of this equation are
\begin{align}
\label{defromed conformal killing vector 1}
\xi^{z}&=\epsilon(z)-\beta\bar \epsilon(\bar z),\\
\label{defromed conformal killing vector 2}
\xi^{\bar z}&=-\alpha\epsilon(z)+\bar \epsilon(\bar z),
\end{align}   
where $\epsilon(z),\bar \epsilon(\bar z)$ are arbitrary functions depend on $z,\bar z$ respectively and $\alpha,\beta$ are introduced for convenient, with 
\begin{align}
\alpha=2L\left(r^2+\frac{L\bar L}{r^2}\right)^{-1},\quad\beta=2\bar L\left(r^2+\frac{L\bar L}{r^2}\right)^{-1}.
\end{align}
Actually, the nonvanishing of $\alpha,\beta$ is the characteristic of $T\bar T$ deformation. As long as we push the boundary to spatial infinity $r\to \infty$, $\alpha,\beta$ tend to vanishing. The infinitesimal transformations become holomorphic and antiholomorphic mapping, which are actually conformal transformations.  
\par
So the general $T\bar T$ deformed conformal killing vectors for BTZ black hole are
\begin{align}
\label{deformed killing vector}
k=\xi^{z}\partial_{z}+\xi^{\bar z}\partial_{\bar z}=\epsilon(z)(\partial_{z}-\alpha\partial_{\bar z})+\bar \epsilon(\bar z)(\partial_{\bar z}-\beta\partial_{z}).
\end{align}
We get the infinitesimal transformation $\xi^{i}=(\xi^{z},\xi^{\bar z})$. According to Noether theorem, there must exist a conserved quantity, which plays an important role in physics. Under this symmetry transformation $x^i\to x^i+\xi^i$, we have a conserved current which can be written as
\begin{align}
j_{i}=T_{ij}\xi^{j}.
\end{align}
It is easy to check this is actually a conserved current because of the trace relation
\begin{align}
\partial^{i}j_{i}&=\partial^{i}T_{ij}\xi^{j}+T_{ij}\partial^{i}\xi^{j}\nonumber\\
&=\partial^{i}T_{ij}\xi^{j}+(T_i^i+2\mu T\bar T)=0. 
\end{align}
The conserved charge is defined as integral on a time slice $\Sigma_t$
\begin{align}
Q=\int_{\Sigma t}\sqrt{g_{\phi\phi}} j_{0}d\phi.
\end{align}
In complex coordinates $z,\bar z$, the conserved charges are
\begin{align}
Q_{\epsilon,\bar\epsilon}=&\int_{\Sigma_t}\sqrt{g_{\phi\phi}}\left[(T_{zz}+T_{z\bar z})-\alpha(T_{z\bar z}+T_{\bar z\bar z})\right]\epsilon(z)d\phi\nonumber\\
&+\int_{\Sigma_t}\sqrt{g_{\phi\phi}}\left[(T_{z\bar z}+T_{\bar z\bar z})-\beta(T_{zz}+T_{z\bar z})\right]\bar\epsilon(\bar z)d\phi,
\end{align} 
where 
\begin{align}
g_{\phi\phi}&=-g_{zz}+2g_{z\bar z}-g_{\bar z\bar z}=1+\frac{L\bar L}{r^4}-\frac{L+\bar L}{r^2},
\end{align}
and on a time slice we have $d\phi=idz=-id\bar z$. The components of stress tensor in $z,\bar z$ coordinate system 
\begin{align}
T_{zz}&=\frac{1}{4}(T_{tt}-T_{\phi\phi}-2iT_{t\phi}),\nonumber\\
T_{\bar z\bar z}&=\frac{1}{4}(T_{tt}-T_{\phi\phi}+2iT_{t\phi}),\nonumber\\
T_{z\bar z}&=\frac{1}{4}(T_{tt}+T_{\phi\phi}).\nonumber
\end{align} 
Then we can get the relations
\begin{align}
T_{zz}+T_{z\bar z}&=\frac{1}{2}(T_{tt}-iT_{t\phi})=\frac{1}{2}(\tau_{tt}-i\tau_{t\phi}),\\
T_{\bar z\bar z}+T_{z\bar z}&=\frac{1}{2}(T_{tt}+iT_{t\phi})=\frac{1}{2}(\tau_{tt}+i\tau_{t\phi}),\\
T_{zz}-T_{\bar z\bar z}&=-iT_{t\phi}=-i\tau_{t\phi}.
\end{align}
In the second step, we used $T_{ij}=\tau_{ij}$, which has been derived in section~\ref{subs: stress tensor}. With these relations, we can express conserved charges as
\begin{align}
Q_{\epsilon,\bar\epsilon}=&\frac{1}{2}\int_{\Sigma_t}\sqrt{g_{\phi\phi}}\left[(1-\alpha)\tau_{tt}-i(1+\alpha)\tau_{t\phi}\right]\epsilon(z)d\phi\nonumber\\
&+\frac{1}{2}\int_{\Sigma_t}\sqrt{g_{\phi\phi}}\left[(1-\beta)\tau_{tt}+i(1+\beta)\tau_{t\phi}\right]\bar\epsilon(\bar z)d\phi.
\end{align} 
\par
The conserved charges can be expressed in terms of spectrum and angular momentum. In fact, according to \eqref{spectrum grav} and \eqref{angular grav}, we can get the formula of $E_n$ and $J_n$ from Brown-York's stress tensor. The normal vector of a time slice in complex coordinates, i.e. $z+\bar z=C$, can be calculated 
\begin{align}
u^{i}=(a,a),\quad a=\frac{r}{\sqrt{r^2+\frac{L\bar L}{r^2}+L+\bar L}}.
\end{align}  
Then the spectrum can be obtained from gravity side 
\begin{align}
E_{n}&=r\int\sqrt{g_{\phi\phi}}\tau_{ij}u^iu^jd\phi=r\int\sqrt{g_{\phi\phi}}a^2(\tau_{zz}+2\tau_{z\bar z}+\tau_{\bar z\bar z})d\phi\nonumber\\
&=ra^2\int\sqrt{g_{\phi\phi}}\tau_{tt}d\phi.
\end{align}  
For angular momentum, which associate with the killing vector $\xi^{i}=\partial_{\phi}$
\begin{align}
\xi^{i}=(b,-b),\quad b=\frac{r}{\sqrt{-(r^2+\frac{L\bar L}{r^2})+L+\bar L}},
\end{align} 
it can be written as  
\begin{align}
J_n&=\int\sqrt{g_{\phi\phi}}\tau_{ij}u^i\xi^jd\phi=\int\sqrt{g_{\phi\phi}}ab(\tau_{zz}-\tau_{\bar z\bar z})d\phi\nonumber\\
&=-abi\int\sqrt{g_{\phi\phi}}\tau_{t\phi}d\phi.
\end{align} 
\par
In case of BTZ black hole, the stress tensor $\tau_{ij}$ and $\sqrt{g_{\phi\phi}}$ do not depend on integration variable $\phi$, with $\phi\sim\phi+2\pi$. Noting the relations above, we can get 
\begin{align}
Q_{\epsilon,\bar\epsilon}
=&\frac{1}{2}\left[\frac{(1-\alpha)E_n}{2\pi ra^2}+\frac{(1+\alpha)J_n}{2\pi ab}\right]\int\epsilon(z)d\phi\nonumber\\
&+\frac{1}{2}\left[\frac{(1-\beta)E_n}{2\pi ra^2}-\frac{(1+\beta)J_n}{2\pi ab}\right]\int\bar\epsilon(\bar z)d\phi. 
\end{align} 
where the $\epsilon(z),\bar\epsilon(\bar z)$ are arbitrary functions and $z,\bar z$ are linearly depend on $\phi$ on a time slice. Following the technique of CFT, we can also perform a mode expansion for the conserved charges. The holomorphic part and antiholomorphic part can be separated by setting $\epsilon=0$ and $\bar\epsilon=0$, respectively 
\begin{align}
Q_{\epsilon}=\sum_{n}\epsilon_{n}Q_n,\quad\bar Q_{\bar \epsilon}=\sum_{m}\bar\epsilon_{m}\bar Q_m,
\end{align}
where
\begin{align}
\label{charge Q}
Q_n=&\frac{1}{2}\left[\frac{(1-\alpha)E_n}{2\pi ra^2}+\frac{(1+\alpha)J_n}{2\pi ab}\right]\int z^{n+1}d\phi,\\
\label{charge Qbar}
\bar Q_m=&\frac{1}{2}\left[\frac{(1-\beta)E_n}{2\pi ra^2}-\frac{(1+\beta)J_n}{2\pi ab}\right]\int\bar z^{m+1}d\phi.
\end{align}
\par
Therefore, we can get the charge for $-1$ mode easily
\begin{align}
Q_{-1}=&\frac{1}{2}\left[\frac{(1-\alpha)E_n}{ra^2}+\frac{(1+\alpha)J_n}{ab}\right], \\
\bar Q_{-1}=&\frac{1}{2}\left[\frac{(1-\beta)E_n}{ra^2}-\frac{(1+\beta)J_n}{ab}\right].
\end{align}
If we take the CFT limit, that is $r\to \infty$, they reduce to 
\begin{align}
Q_{-1}\to\frac{1}{2}(M+J)=L,\\
\bar Q_{-1}\to\frac{1}{2}(M-J)=\bar L.
\end{align}
which coincide with the CFT case marked as $L_0,\bar L_0$~\cite{Banados:1992gq, banados1999three}. The $-1$ mode corresponds to zero mode of CFT, because the radial quantization is used  in CFT and time translation becomes dilation. 
\par
The charges for other modes can also be calculated from \eqref{charge Q}, \eqref{charge Qbar}, which are proportional to $Q_{-1}$ or $\bar Q_{-1}$. So we can not get an infinite amount of independent conserved charges. The reason might be that we can only obtain the global charges by this method, or the aspects of BTZ black hole whose stress tensor does not depend on $\phi$. Actually, BTZ black hole has only two charges which are non-zero $M=L_0+\bar L_0,\ J=L_0-\bar L_0$. However, $T\bar T$ deformed CFT as an integrable system, it is still not much clear about the infinite amount of conserved charges, even though we have a symmetry.         
\section{Conclusion and discussion}
\label{sec: con}
The purpose of this paper is to explore the symmetry of $T\bar T$ deformed CFT. We are based on the idea that $T\bar T$ flow may turn the conformal symmetry to others, because this integrable deformation preserves an infinite amount of conserved charges. We restudy the $T\bar T$ deformed CFT, and find the traceless of CFT becomes a trace relation along the $T\bar T$ flow. The former one implies the conformal symmetry. Naturally, we get the symmetry by replacing the traceless with trace relation. The deformed conformal killing equation is obtained, which is not only a flow equation but also implies a symmetry. Moreover, the proposal that $T\bar T$ deformed CFT corresponds to a cutoff AdS, inspired us to consider the boundary theory placing on a dynamical background. The deformation term is just the variation of metric along parameter $\mu$, and conformal symmetry coinsides with the fixed point. From a holographic perspective, this symmetry is an asymptotic symmetry of AdS$_3$ with a new boundary condition, which includes higher order in Fefferman-Graham gauge. Brown-Henneaux's boundary condition just preserves the leading order and allows an asymptotic symmetry of induced metric in this case. Therefore this symmetry is a generalization of asymptotic behavior of AdS$_3$ with high order correction. However, we just discuss this symmetry on a classical level. Besides, the deformed conformal killing equation is not covariant, which is coordinate dependent. On the gravity side, the conclusion relies on Fefferman-Graham gauge.
\par     
As a result of the deformed conformal killing equation refer to the stress tensor of field theory, it provides an approach to calculate the stress tensor from gravity side. Actually, we argued the holographic duality of stress tensor, that is the stress tensor of $T\bar T$ equals to Brown-York's quasilocal stress tensor with a counterterm.
\par
In order to obtain a specific symmetry, we consider BTZ black hole. The $T\bar T$ deformed conformal killing equation gives the infinitesimal transformations. We also computed the conserved charges of this symmetry. The result is turned to be consistent with the CFT case, when we push the boundary to spatial infinity. However, we still cannot get an infinite set of conserved charges. The reason may be the trace relation does not include all symmetry of this system or maybe because of the special feature of BTZ black hole. All in all, we did not get an infinite amount of independent conserved charges. So we can not get a nontrivial symmetry algebra of conserved charges about $T\bar T$ deformed CFT by this method. 
\par
However, analogous to the conformal symmetry, we can get an infinite set of generators about the  infinitesimal symmetry transformation from \eqref{deformed killing vector} by performing a modes expansion with respect to $\epsilon(z),\bar\epsilon(\bar z)$. These generators can constitute an algebra formally. In fact, we can write down the generators 
\begin{align}
L_n&=-z^{n+1}(\partial_z-\alpha\partial_{\bar z})=l_n-\alpha j_n,\\
\bar L_m&=-\bar z^{m+1}(\partial_{\bar z}-\beta\partial_{z})=\bar l_m-\beta\bar j_m,
\end{align}   
Here, $l_n,\bar l_m$ are the familiar conformal generators, which obey the commutation relations
\begin{align}
[l_n,l_m]=(n-m)l_{n+m},[l_n,\bar l_m]=0, [\bar l_n,\bar l_m]=(n-m)\bar l_{n+m}.
\end{align}
After performing a nontrivial central extension, it becomes Virasoro algebra with a central charge. Something new about $T\bar T$ deformed CFT is the presence of $j_n,\bar j_m$.
These yield the commutation relations of $T\bar T$ symmetry generators $L_n,\bar L_m$
\begin{align}
[L_n,L_m]=&(n-m)L_{n+m},\\ [\bar L_n,\bar L_m]=&(n-m)\bar L_{n+m},\\
[L_n,\bar L_m]=&-\beta(n+1)\bar z^{m+1}L_{n-1}+\alpha(m+1)z^{n+1}\bar L_{m-1},
\end{align}   
from which, one can see that these generators form an algebra. Up to a central extension, the difference between this algebra and Virasoro algebra is the commutation relation $[L_n,\bar L_m]$, which is vanishing in case of the later one. Of course, the generators $L_n,\bar L_m$ are also different from conformal generator by definition, because the nonvanishing of $\alpha,\beta$. If we push the boundary to infinity, this algebra reduces to Virasoro algebra. From an algebraic viewpoint, $T\bar T$ deformation effect is coupling the two copies of Virasoro algebra. But we should point out that the conserved charges associate with these generators are not independent, which we have calculated in Section~\ref{sec: BTZ}. So we still do not clear about this symmetry algebra in a deep level. 
\par
In addition, AdS$_3$ gravity can be formulated as Chern-Simons theory, and gauge transformations can produce the diffeomorphisms~\cite{Witten:1988hc}. Global charges in Chern-Simons theory are related to a particular Virasoro algebra via a twisted Sugawara construction $\mathcal L_{\xi} g_{\phi\phi}=0$ in ~\cite{Banados:1994tn}, see also~\cite{, Banados:1998ta, Carlip:1998qw}.
Therefore, Chern-Simons theory may provide a route to study the symmetry and holographic aspects of $T\bar T$ deformed CFT. In particular, with the boundary condition of $T\bar T$, namely $\mathcal L_{\xi}\gamma_{ij}=0$, we have the diffeomorphisms in \eqref{defromed conformal killing vector 1} and \eqref{defromed conformal killing vector 2}. In terms of Chern-Simons formula, the global charges can be calculated 
\begin{align}
\mathcal Q_0=&(1-\beta)L,\\
\bar {\mathcal Q}_0=&(1-\alpha)\bar L,
\end{align}
which are proportional to $L,\bar L$. But the effect of $T\bar T$ deformation makes these different from $L,\bar L$ because of non-vanishing of $\alpha,\beta$. For the CFT case $\alpha,\beta\to 0$, these charges reduce to $L,\bar L$. However, the charges are different from the $Q_{-1},\bar Q_{-1}$ at finite radial coordinate $r$, because they are defined by different ways. The later are quasilocal conserved charges defined by Hamilton-Jacobi method from gravity side, which also have holographic aspects. But global charges in Chern-Simons are defined by surface integral method in a gauge theory. Whether or not, it is a way to research the $T\bar T$ deformed CFT. The detail calculation and discussion about Chern-Simons formalism are affixed to the Appendix~\ref{cs}. It would be interesting to understand and explore more about the $T\bar T$ deformed CFT and holographic properties from different theories.
\section*{Acknowledgments}
We would like to thank Feng Qu for useful discussions. This work is supported by the National Natural Science Foundation of China (NSFC) with Grant No.11847612, No.11875082.
\appendix 
\section{Chern-Simons formalism}
\label{cs}
In this appendix, we treat in more detail to calculate the charges from Chern-Simons formalism. We start from writing AdS$_3$ in gauge fields form. Three dimensional Einstein gravity with a negative cosmological constant can be expressed as $SL(2,\mathbb{R})\times SL(2,\mathbb{R})$ Chern-Simons gauge theory, with the action
\begin{align}
S_{grav}=I(A)-I(\bar A),
\end{align} 
where 
\begin{align}
I(A)=\frac{\kappa}{4\pi}\int_{M}\text{Tr}\left(A\wedge dA+\frac{2}{3}A\wedge A \wedge A \right),\kappa=\frac{l}{4G}
\end{align} 
and 
\begin{align}
A^a=\omega^a+\frac{1}{l}e^a, \bar A^a=\omega^a-\frac{1}{l}e^a.
\end{align}
In terms of triads, the metric \eqref{btz metric} can be formulated as gauge fields
\begin{align}
A=&\left(\begin{matrix}
	dr/2r & Ldz/r\\
	rdz & -dr/2r\\
\end{matrix}\right)=(r-\frac{L}{r})J_0dz+\frac{1}{r}J_1dr+(r+\frac{L}{r})J_2dz,\\
\bar A=&\left(\begin{matrix}
	-dr/2r & rd\bar z\\
	\bar Ld\bar z/r & dr/2r\\
\end{matrix}\right)=-(r-\frac{\bar L}{r})J_0d\bar z-\frac{1}{r}J_1dr+(r+\frac{\bar L}{r})J_2d\bar z,
\end{align}
where the bases $J_a$ are generators of $SL(2,\mathbb{R})$
\begin{align}
J_0=\frac{1}{2}\left(
  \begin{matrix}
   0 & -1 \\
   1 & 0 
  \end{matrix}
  \right),
J_1=\frac{1}{2}\left(
  \begin{matrix}
   1 & 0 \\
   0 & -1 
  \end{matrix}
  \right),
J_2=\frac{1}{2}\left(
  \begin{matrix}
   0 & \ 1 \\
   1 & \ 0 
  \end{matrix}
  \right),
\end{align}
with commutation relations and killing metric 
\begin{align}
[J_a,J_b]=\varepsilon_{ab}^{\ \ c}J_c,\qquad g_{ab}=\text{Tr}(J_aJ_b)=\frac{1}{2}\eta_{ab}=\frac{1}{2}\text{diag}(-1,1,1).
\end{align}
Global charges in Chern-Simons theory are defined as 
\begin{align}
{\mathcal Q}(\eta)=\frac{\kappa}{4\pi}\int_{\partial\Sigma}\text{Tr}\left(\eta A\right),
\end{align}
where the parameter of gauge transformation $\eta$ is related to the diffeomorphisms $\xi^{i}$ by
\begin{align}
\eta^a=\xi^iA_i^a.
\end{align} 
In ~\cite{Banados:1998ta, Carlip:1998qw, Banados:1994tn}, Ba\~nados and Carlip reproduced the Virasoro algebra from the Poisson bracket of the global charges by setting the boundary condition $\mathcal L_{\xi}g_{\phi\phi}=0$. That is the twisted Sugawara construction $\xi^r=-C(r)\partial_{\phi}\xi^{\phi}$, and $\xi^{\phi}$ just depends on $\phi$. In Schwarzschild coordinates, we have $C(r)=r/lN(r)$, where $N(r)$ is lapse function of BTZ black hole. After a partial integral, the charges can be expressed as 
\begin{align}
{\mathcal Q}(\eta)&=\frac{\kappa}{4\pi}\int_{\partial\Sigma}g_{ab}\eta^aA^b_{\phi}d\phi=\frac{\kappa}{4\pi}\int_{\partial\Sigma}g_{ab}(2C(r)\xi^{\phi}A_r^a\partial_{\phi}A^b_{\phi}+\xi^{\phi}A^a_{\phi}A^b_{\phi})d\phi.
\end{align}
But in our coordinates, we make a radial decomposition and treat the radial as time. As the gauge fields do not depend on $z,\bar z$, we can get the charges easily in case of BTZ black hole
\begin{align}
{\mathcal Q}(\xi)&=\frac{\kappa}{4\pi}\int_{\partial\Sigma_t}g_{ab}\xi^{z}A^a_{z}A^b_{z}d\phi=\frac{\kappa}{2\pi}\int_{\partial\Sigma_t}\xi^zLd\phi.
\end{align}
Similarly, we can get 
\begin{align}
\bar {\mathcal Q}(\bar\xi)&=\frac{\kappa}{4\pi}\int_{\partial\Sigma_t}g_{ab}\xi^{\bar z}\bar A^a_{\bar z}\bar A^b_{\bar z}d\phi=\frac{\kappa}{2\pi}\int_{\partial\Sigma_t}\xi^{\bar z}\bar Ld\phi.
\end{align}
We can also put it in terms of Fourier modes. The non-zero charges are zero mode, which related to $M$ and $J$ by $Q_0=L\equiv(M+J)/2,\bar Q_0=\bar L\equiv(M-J)/2$ (up to an unessential coefficient). These charges are also hold for the boundary at finite $r$, because the boundary condition gives $\xi^r=-C(r)\partial_{\phi}\xi^{\phi}$ that holds for all radial coordinate. But $C(r)$ would lead to a shift in central charge of the Virasoro algebra for different $r$. The definition of charge in Chern-Simons inspires us to consider the charge of $T\bar T$ deformed CFT.
\par
Now we will put the $T\bar T$ boundary condition of the induced metric $\gamma_{ij}$ of \eqref{boundary condition} in this formula. This boundary condition is stronger than $\mathcal L_{\xi}g_{\phi\phi}=0$ because it restricts the whole induced metric not only $g_{\phi\phi}$ component. We have got the solution of $\mathcal L_{\xi}\gamma_{ij}=0$ exactly, i.e. \eqref{defromed conformal killing vector 1} and \eqref{defromed conformal killing vector 2}. Then the charges in Chern-Simons theory are 
\begin{align}
{\mathcal Q}(\xi^i)=&\frac{\kappa}{4\pi}\int_{\partial\Sigma_t}g_{ab}(\xi^{\bar z}A^a_{\bar z}+\xi^{z}A^a_{z})A^b_zd\phi.
\end{align} 
Noting the relations
\begin{align}
A^a_{\bar z}=0,\quad\bar A^a_{z}=0,\quad g_{ab}A^a_{z}A^b_z=2L,\quad g_{ab}\bar A^a_{\bar z}\bar A^b_{\bar z}=2\bar L,
\end{align} 
and conventions we used in section 5, we arrive at
\begin{align}
\mathcal Q(\xi^i)=&\frac{\kappa}{2\pi}\int_{\partial\Sigma_t}[\epsilon(z)-\beta\bar \epsilon(\bar z)]Ld\phi,\\
\bar {\mathcal Q}(\xi^i)=&\frac{\kappa}{2\pi}\int_{\partial\Sigma_t}[-\alpha\epsilon(z)+\bar \epsilon(\bar z)]\bar Ld\phi.
\end{align}
Here $d\phi=idz=-id\bar z$ on a time slice $\Sigma_t$, so we just write these as $d\phi$. The charges are proportional 
to $L,\bar L$. Modes expansion of $\epsilon(z),\bar\epsilon(\bar z)$ allow us to write down the zero mode which can be compared with CFT case
\begin{align}
\mathcal Q_0=&(1-\beta)L,\\
\bar {\mathcal Q}_0=&(1-\alpha)\bar L.
\end{align}
These charges are associated with the boundary condition of $T\bar T$ deformation, which is different from the case of Ba\~nados at finite $r$ in~\cite{Banados:1998ta}, because of $\alpha,\beta$. Pushing the $T\bar T$ symmetry to spatial infinity, $\alpha,\beta$ tend to vanishing, and the charges reduce to the CFT case. 
\par
\providecommand{\href}[2]{#2}\begingroup\raggedright\endgroup

\end{document}